\documentstyle[preprint,aps]{revtex}
\begin{document}

\draft

\title{Tunneling Conductance and Coulomb Blockade Peak Splitting of Two
Quantum Dots Connected by a Quantum Point Contact}

\author{Yu-Liang Liu}
\address{Max-Planck-Institut f\"{u}r Physik Komplexer Systeme, Bayreuther
Str. 40, D-01187 Dresden, Germany}

\maketitle

\begin{abstract}

By using bosonization method and unitary transformation, we give a general
relation between the dimensionless tunneling conductance and the fractional
Coulomb blockade conductance peak splitting which is valid both for weak and
strong transmission between two quantum dots, and show that the tunneling
conductance has a linear temperature dependence in the low energy and
low temperature limit.

\end{abstract}
\vspace{1cm}


\newpage

Electron tunneling in a mesoscopic structure may be
drastically influenced by the
charging effects. If the
charge spreading is impeded by weak links, or by a special geometry of the
structure, the charging suppresses the electron tunneling. Such a suppression
of the electron tunneling is commonly called as the Coulomb
blockade\cite{1,2,3}. It has become possible to observe the Coulomb blockade
effect in semiconductor heterostructures where the geometry of the system can
be easily modified by adjusting the voltages on special gate electrodes. 
Recently, the electron tunneling through two quantum dots connected by single
quantum point contact which is controlled by adjusting voltage on additional
gate electrode
has been extensively studied, both experimental\cite{4,5,6,7,8} 
and theoretical\cite{9,10,11,12,13}. 
The electron tunneling between two quantum dots leads
to a decay of the Coulomb blockade of the individual dot.
For a pair of electrostatically identical quantum dots, the progress of this
decay can be described by tracking the splitting of the Coulomb blockade
conductance peaks as they evolve from doubly degenerate single-dot conductance
resonances to that of nondegenerate double-dot peaks with twice the original
periodicity. 
The Coulomb blockade peak splitting significantly depends upon the
transmission coefficient ${\cal T}$ of the quantum point contact, i.e., the
tunneling conductance. For the case of ${\cal T}=1$, the Coulomb blockade
between two quantum dots
disappears, they become
a double-quantum-dot. For the case of ${\cal T}=0$, the Coulomb blockade is
maximum which completely suppresses the electron tunneling between two quantum
dots, therefore, they are separated from each other. 

In this paper, by using bosonization method and unitary transformation
developed in Ref.\cite{14}, we can effectively treat the system of two
electrostaticaly identical quantum dots connected by one quantum point
contact which is controlled by adjusting voltage on additional gate electrode,
and give a general relation between the dimensionless tunneling conductance $g$
and the fractional Coulomb blockade peak splitting $f$, which is
valid both for weak (${\cal T}\sim 0$) and strong (${\cal T}\sim 1$)
transmission coefficients. We also study the temperature dependence of the
tunneling conductance in the low energy and low temperature limit. 

In generally, each individual quantum dot can be described by two-dimensional
electron gas, and the Coulomb charging energy $E_{c}=e^{2}/(2C)$, where $C$ is
the capacitance of the individual quantum dot, is very large compared to the
single-particle level spacing but small compared to the tunneling channel
band-widths. For the case of only a few modes can tunnel from one quantum dot
to another one, the quantum point contact can be described by a quantum
wire\cite{15,16}. 
We can also use a potential barrier to adjust the transmission
coefficient.  Based upon above considerations, we consider the following
Hamiltonians\cite{15,16,17} 
\begin{equation}
H_{0}=-i\hbar v_{F}\sum_{\sigma}
\int dx [\psi^{+}_{R\sigma}(x)\partial_{x}\psi_{R\sigma}(x)
-\psi^{+}_{L\sigma}(x)\partial_{x}\psi_{L\sigma}(x)]
\label{1}\end{equation}
\begin{equation}
H_{c}=E_{c}(\hat{n}-\frac{N}{2})^{2}
\label{2}\end{equation}
\begin{equation}
H_{I}=V_{2k_{f}}\sum_{\sigma}[\psi^{+}_{R\sigma}(0)\psi_{L\sigma}(0)
+\psi^{+}_{L\sigma}(0)\psi_{R\sigma}(0)]
\label{3}\end{equation}
where $\psi_{R\sigma}(x)$ are right-moving electron operators, 
$\psi_{L\sigma}(x)$ are left-moving electron operators;
$E_{c}=e^{2}/(2C)$, $eN=C_{g}(V_{gR}-V_{gL})$ is the gate voltage parameter 
of two quantum dots, and $C_{g}$ is a gate-to-dot capacitance.
$e\hat{n}=(e/2)\sum_{\sigma}
\int^{\infty}_{0} dx[\psi^{+}_{R\sigma}(x)
\psi_{R\sigma}(x)-\psi^{+}_{L\sigma}(-x)\psi_{L\sigma}(-x)]$ is the charge
difference between two quantum dots; $V_{2k_{F}}$ is a backward scattering
potential which controls the transmission
coefficient ${\cal T}$. 
The system described by these Hamiltonians (\ref{1}), (\ref{2}) and (\ref{3})
has been extensively studied by directly using the bosonization representation
of the electron fields $\psi_{R(L)\sigma}(x)$ in the strategy of perturbation
methods. However, the backward scattering term is relevant in the terminology
of the renormalization group, the perturbation method may fail. To more
effectively study this problem, we adopt another way developed in Ref.\cite{14}
where the backward scattering term can be exactly cancelled by an unitary
transformation, and its effect is reflected on the correlation functions of
the electrons and the expression of the charge operator $e\hat{n}$. We 
define the following new fermion fields
\begin{equation}
\psi_{1\sigma}(x)=\frac{1}{\sqrt{2}}[\psi_{R\sigma}(x)+\psi_{L\sigma}(-x)],
\;\;\; \psi_{2\sigma}(x)=\frac{1}{\sqrt{2}}
[\psi_{R\sigma}(x)-\psi_{L\sigma}(-x)]
\label{4}\end{equation}
while the bosonic representation of these fermion fields can be written  
in the usual way\cite{18,19,20} as
\begin{equation}
\psi_{1(2)\sigma}(x)=(\frac{D}{2\pi\hbar v_{F}})^{1/2}
e^{-i\Phi_{1(2)\sigma}(x)}
\label{5}\end{equation}
where $D$ is the band width of the conduction electrons. The boson fields
$\Phi_{1(2)\sigma}(x)$ have the relation with the density operators
$\partial_{x}\Phi_{1(2)\sigma}(x)=2\pi\rho_{1(2)\sigma}(x)$,
and satisfy the commutation relations $[\partial_{x}\Phi_{1\sigma}(x),\;
\Phi_{1\sigma^{'}}(y)]=i2\pi\delta_{\sigma\sigma^{'}}\delta(x-y)$,
$[\partial_{x}\Phi_{2\sigma}(x),\;
\Phi_{2\sigma^{'}}(y)]=i2\pi\delta_{\sigma\sigma^{'}}\delta(x-y)$,
where the density operators are defined as $\rho_{1(2)\sigma}(x)=
\psi^{+}_{1(2)\sigma}(x)\psi_{1(2)\sigma}(x)$. The Hamiltonians (\ref{1}) and
(\ref{3}) can be written in the bosonic representation of the fermion
fields $\psi_{1(2)\sigma}(x)$ as
\begin{equation}
H_{0}=\frac{\hbar v_{F}}{4\pi}\sum_{\sigma}\int dx[
(\partial_{x}\Phi_{1\sigma})^{2}+(\partial_{x}\Phi_{2\sigma})^{2}]
\label{6}\end{equation}
\begin{equation}
H_{I}=\frac{\hbar v_{F}\delta}{2\pi}\sum_{\sigma}[
\partial_{x}\Phi_{1\sigma}(x)-\partial_{x}\Phi_{2\sigma}(x)]|_{x=0}
\label{7}\end{equation}
where $\delta=\arctan(V_{2k_{F}}/(\hbar v_{F}))$ is phase shift induced by the
backward scattering potential. According to
Eq.(\ref{4}), the charge operator $e\hat{n}$ can be written as
\begin{equation}
\hat{n}=\frac{1}{2}\sum_{\sigma}\int^{\infty}_{0}\!\! dx[
\psi^{+}_{1\sigma}(x)\psi_{2\sigma}(x)+\psi^{+}_{2\sigma}(x)\psi_{1\sigma}(x)
-\psi^{+}_{1\sigma}(-x)\psi_{2\sigma}(-x)-
\psi^{+}_{2\sigma}(-x)\psi_{1\sigma}(-x)]
\label{8}\end{equation}
which only includes the cross terms of the fermion fields
$\psi_{1\sigma}(x)$ and $\psi_{2\sigma}(x)$.

In order to cancel the $\delta$-term in (\ref{7}), we define the following 
unitary transformation
\begin{equation}
U=\exp\{i\frac{\delta}{2\pi}\sum_{\sigma}[\Phi_{1\sigma}(0)
-\Phi_{2\sigma}(0)]\}
\label{9}\end{equation}
Under this unitary
transformation, we can obtain the following relations
\begin{eqnarray}
U^{+}(H_{0}+H_{I})U &=& H_{0} \nonumber \\
\bar{H}_{c}=U^{+}H_{c}U &=& \displaystyle{E_{c}[U^{+}\hat{n}U-\frac{N}{2}]^{2}}
\label{10}\end{eqnarray}
It is worth notice that the backward scattering term completely disappears,
its effect is only reflected on the charge operator $e\hat{n}$ for this
problem. It has not any influence on the correlation functions of the fermion
fields $\psi_{1(2)\sigma}(x)$ because we do not consider the electron-electron
interactions in the quantum wire.
The calculation of $U^{+}\hat{n}U$ is very simple. By using the following
formulae 
\begin{eqnarray}
[\Phi_{1\sigma}(x), \; \Phi_{1\sigma^{'}}(y)] &=& i\pi\delta_{\sigma\sigma'}
sgn(x-y), \;\;\;
[\Phi_{2\sigma}(x), \; \Phi_{2\sigma^{'}}(y)] = i\pi\delta_{\sigma\sigma'}
sgn(x-y) \nonumber \\ 
U^{+}\Phi_{1\sigma}(x)U &=& \Phi_{1\sigma}(x) + \frac{\delta}{2} sgn(x),
\;\;\; U^{+}\Phi_{2\sigma}(x)U=\Phi_{2\sigma}(x)-\frac{\delta}{2} sgn(x)
\label{11}\end{eqnarray}
and taking the following gauge transformations
\begin{equation}
\psi_{1\sigma}(x)=\bar{\psi}_{1\sigma}(x)e^{i\theta_{1}},\;\;
\psi_{2\sigma}(x)=\bar{\psi}_{2\sigma}(x)e^{i\theta_{2}},\;\;
\theta_{1}-\theta_{2}=\delta
\label{12}\end{equation}
we can easily obtain the following expression of the charge operator
$eU^{+}\hat{n}U$ 
\begin{eqnarray}
U^{+}\hat{n}U &=& \hat{n}+\displaystyle{\frac{\cos(2\delta)-1}{2}
\sum_{\sigma}\int dx[
\bar{\psi}^{+}_{1\sigma}(x)\bar{\psi}_{2\sigma}(x)+
\bar{\psi}^{+}_{2\sigma}(x)\bar{\psi}_{1\sigma}(x)]} \nonumber \\
&-&i\displaystyle{\frac{\sin(2\delta)}{2}\sum_{\sigma}\int dx[
\bar{\psi}^{+}_{1\sigma}(x)\bar{\psi}_{2\sigma}(x)-
\bar{\psi}^{+}_{2\sigma}(x)\bar{\psi}_{1\sigma}(x)]}
\label{13}\end{eqnarray}
It is noted that although the backward scattering term is eliminated by the
unitary transformation $U$ (\ref{9}), the charge operator $eU^{+}\hat{n}U$
becomes complex. To simplify it,
we re-define the right- and left-moving electronic fields
\begin{equation}
\bar{\psi}_{R\sigma}(x)=\frac{1}{\sqrt{2}}[\bar{\psi}_{1\sigma}(x)+
\bar{\psi}_{2\sigma}(x)],\;\;\;
\bar{\psi}_{L\sigma}(x)=\frac{1}{\sqrt{2}}[\bar{\psi}_{1\sigma}(-x)-
\bar{\psi}_{2\sigma}(-x)]
\label{14}\end{equation}
while their bosonic representations reads $\bar{\psi}_{R(L)\sigma}(x)=
(\frac{D}{2\pi\hbar v_{F}})^{1/2} \exp\{-i\bar{\Phi}_{R(L)\sigma}(x)\}$. 
These new right- and left-moving electronic fields are
different from original ones due to we have taken the unitary and gauge
transformations of the fermion fields $\psi_{1(2)\sigma}(x)$. In terms of 
these new electronic fields, the charge operator
$eU^{+}\hat{n}U$ can be rewritten as
\begin{eqnarray}
U^{+}\hat{n}U &=& \displaystyle{n_{\delta}+
\frac{\cos(2\delta)+1}{4\pi}\sum_{\sigma}
(\bar{\Phi}_{L\sigma}(0)-\bar{\Phi}_{R\sigma}(0))} \nonumber \\
&-& i \displaystyle{\frac{\sin(2\delta)}{2}\sum_{\sigma}\int^{\infty}_{0}dx[
\bar{\psi}^{+}_{L\sigma}(-x)\bar{\psi}_{R\sigma}(x)
-\bar{\psi}^{+}_{R\sigma}(x)\bar{\psi}_{L\sigma}(-x)]}
\label{15}\end{eqnarray}
where $n_{\delta}=(1/(4\pi))\sum_{\sigma}[
\cos(2\delta)\bar{\phi}_{R\sigma}(\infty)+\bar{\phi}_{R\sigma}(-\infty) 
-\bar{\phi}_{L\sigma}(\infty) 
-\cos(2\delta)\bar{\phi}_{L\sigma}(-\infty)]$.
If we take the value of the phase shift $\delta$ as $\delta^{c}=\pm\pi/2$, we
can obtain the following relation
\begin{equation}
U^{+}\hat{n}U|_{\delta^{c}}=n_{\delta^{c}}=\frac{1}{2}\sum_{\sigma}
(\bar{N}_{L\sigma}-\bar{N}_{R\sigma})
\label{16}\end{equation}
where $\bar{N}_{R(L)\sigma}=\int^{\infty}_{-\infty}dx
\bar{\psi}^{+}_{R(L)\sigma}(x)\bar{\psi}_{R(L)\sigma}(x)$ are the right- and
left-moving electron numbers, respectively. 
Therefore, the value of the phase shift
$\delta^{c}=\pm\pi/2$ corresponds to strong coupling critical point of the
system induced by the backward scattering potential. At
this strong coupling critical point, the transmission coefficient ${\cal T}$ of
the electrons from the left quantum dot to the right one (or vice versa) is
zero, i.e., the electrons are completely reflected on the impurity site $x=0$.
These two quantum dots are completely separated, therefore, 
we can observe a series
of the double degenerate Coulomb blockade conductance peaks at this strong
coupling critical point. 

We now study the low temperature behavior of the tunneling conductance
and the fractional Coulomb blockade conductance peak splitting away from this
strong coupling critical point ($\delta^{c}=\pm\pi/2$). Because the directly
hopping of the electrons from the right of the impurity to its left or vice
versa is very weak, we can approximately replace the quantity
\[
\int^{\infty}_{0}dx[\bar{\psi}^{+}_{L\sigma}(-x)\bar{\psi}_{R\sigma}(x)-
\bar{\psi}^{+}_{R\sigma}(x)\bar{\psi}_{L\sigma}(-x)]
\]
by 
\[
a[\bar{\psi}^{+}_{L\sigma}(0)\bar{\psi}_{R\sigma}(0)-
\bar{\psi}^{+}_{R\sigma}(0)\bar{\psi}_{L\sigma}(0)]
\]
where $a$ is a small constant. For simplicity, we define the following new
boson fields
\[
\Phi_{\pm c}(x) = \displaystyle{\frac{1}{2}[
\bar{\Phi}_{L\uparrow}(x)+\bar{\Phi}_{L\downarrow}(x)\pm
(\bar{\Phi}_{R\uparrow}(x)+
\bar{\Phi}_{R\downarrow}(x))]}
\]
\[
\Phi_{\pm s}(x) = \displaystyle{\frac{1}{2}[
\bar{\Phi}_{L\uparrow}(x)-\bar{\Phi}_{L\downarrow}(x)\pm
(\bar{\Phi}_{R\uparrow}(x)-
\bar{\Phi}_{R\downarrow}(x))]}
\]
In terms of these new boson fields, 
the charge operator $eU^{+}\hat{n}U$ is
\begin{equation}
U^{+}\hat{n}U = n_{\delta}+\frac{\cos(2\delta)+1}{2}
\Phi_{-c}(0)+\frac{aD\sin(2\delta)}{\pi\hbar v_{F}}\sin(\Phi_{-c}(0))
\cos(\Phi_{-s}(0))
\label{17}\end{equation}
and the Hamiltonian (\ref{6}) reads
\begin{equation}
H_{0}=\frac{\hbar v_{F}}{4\pi}\int dx[
(\partial_{x}\Phi_{+c}(x))^{2}+(\partial_{x}\Phi_{+s}(x))^{2}+
(\partial_{x}\Phi_{-c}(x))^{2}+(\partial_{x}\Phi_{-s}(x))^{2}]
\label{18}\end{equation}
However, due to the quantity $a\sin(2\delta)$ is very small, the last term in
(\ref{17}) can be perturbatively treated, the electrostatic Hamiltonian
(\ref{10}) can be written as by using Eq.(\ref{17})
\begin{equation}
\bar{H}_{c}=E_{c}[\frac{\cos(2\delta)+1}{2\pi}\Phi_{-c}(0)-\Delta N]^{2}
+E_{c}(\frac{aD\sin(2\delta)}{\pi\hbar v_{F}})^{2}
\sin^{2}(\Phi_{-c}(0))\cos^{2}(\Phi_{-s}(0))
\label{19}\end{equation}
where $\Delta N=-n_{\delta}+N/2$. It is worth noting that for the case of 
weak backward scattering $\delta\sim 0$, 
i.e., the transmission coefficient ${\cal T}\sim 1$,
the low temperature behavior of the system is different from that for the
strong backward scattering. We first consider
the case of the weak backward scattering $\delta\sim 0$. Because the factor $
(\cos(2\delta)+1)/(2\pi)$ is finite, the boson field $\Phi_{-c}(0)$ has a
non-zero expectation value
\begin{equation}
\frac{\cos(2\delta)+1}{2\pi}<\Phi_{-c}(0)>=\Delta N
\label{20}\end{equation}
which suppresses its low energy excitations. 
Therefore, we can safely integrate out the boson field
$\Phi_{-c}(0)$ and obtain the following effective Hamiltonian
\begin{equation}
\bar{H}^{'}_{c}=E_{c}(\frac{aD\sin(2\delta)}{2\pi\hbar v_{F}})^{2}
[1-(\frac{4\pi e^{\gamma}\bar{E}_{c}}{D})^{2}\cos(2<\Phi_{-c}(0)>)]
[1+\cos(2\Phi_{-s}(0))]
\label{21}\end{equation}
where $\bar{E}_{c}=E_{c}(1+\cos(2\delta))^{2}/(2\pi)^{2}$, and $\gamma\simeq
0.577$ is Euler constant. However, in the fermionization representation,
$\cos(2\Phi_{-s}(0))$ can be written as: $\cos(2\Phi_{-s}(0))\sim
\psi^{+}(0^{+})\psi^{+}(0^{-})+\psi(0^{-})\psi(0^{+})$, where
$0^{\pm}=\pm\rho$, $\rho$ is an infinitesimal quantity, which produces a gap 
in the spectrum of fermion $\psi(0)$. Therefore, the
$\cos(2\Phi_{-s}(0))$-term in (\ref{21}) is irrelevant in the terminology of
the renormalization group. 

For the case of the strong backward scattering $\delta\sim\pm\pi/2$, the
electrostatic energy significantly depends upon the quantity $\Delta N$. If 
$\Delta N$ is zero, $\Delta N=0$, after integrating out 
the boson field $\Phi_{-c}(0)$, we
can obtain the following electrostatic Hamiltonian
\begin{equation}
\bar{H}^{'}_{c}=E_{c}(\frac{aD\sin(2\delta)}{2\pi\hbar v_{F}})^{2}
[1-(\frac{4\pi e^{\gamma}\bar{E}_{c}}{D})^{2}]
[1+\cos(2\Phi_{-s}(0))]
\label{22}\end{equation}
For the case of $\Delta N\neq 0$, because of $|Max(\Phi_{-c}(0))|=\pi$, 
if the phase shift $\delta$ satisfies the
relation: $|\delta |\leq\delta_{0}$, where $\delta_{0}$ is defined as:
$1+\cos(2\delta_{0})=2\Delta N$, after integrating out the boson field 
$\Phi_{-c}(0)$, we can obtain the same Hamiltonian as that in (\ref{21}).
If the phase shift satisfies the relation: $\delta_{0}< |\delta|\leq \pi/2$,
we can obtain the following effective Hamiltonian
\begin{eqnarray}
\bar{H}^{'}_{c} &=& \displaystyle{E_{c}[\frac{\cos(2\delta)+1}{2\pi}
<\Phi_{-c}(0)>-\Delta N]^{2}} \nonumber \\
&+& \displaystyle{E_{c}(\frac{aD\sin(2\delta)}{2\pi\hbar v_{F}})^{2}
[1-(\frac{4\pi e^{\gamma}\bar{E}_{c}}{D})^{2}]
[1+\cos(2\Phi_{-s}(0))]}
\label{23}\end{eqnarray}
where $<\Phi_{-c}(0)>=\pi$ for $\Delta N>0$, and 
$<\Phi_{-c}(0)>=-\pi$ for $\Delta N<0$. 
It is worth notice that the present method used to get the final effective
Hamiltonians in (\ref{21}), (\ref{22}) and (\ref{23}) is completely 
different from that in Refs.\cite{15} and \cite{21}.
After performing the unitary transformation (9), all information of the 
conduction electron
scattering on the impurity is incorporated into the electrostatic 
Hamiltonian (19)
which only contains higher order terms of the boson fields $\Phi_{-c(s)}(0)$.
However, in Refs.\cite{15} and \cite{21}, the authors directly take the
approximation on the backward scattering potential which leads to a lower
order terms of the boson fields $\Phi_{-c(s)}(0)$. This difference may make us
obtain a different transport behavior from that
in Ref.\cite{21} in the low temperature limit. On the other hands, in terms of
the terminology of the renormalization group, the former only gives some
irrelevant terms, but the latter gives some relevant terms.
Based upon above discussions, for the
case of $0\leq |\delta|\leq\delta_{0}$, we can obtain the following ground
state energy of the system
\begin{equation}
E=E_{0}-E_{c}(\frac{2ae^{\gamma}\bar{E}_{c}\sin(2\delta)}{\hbar v_{F}})^{2}
\cos(2<\Phi_{-c}(0)>)
\label{24}\end{equation}
For the case of $\delta_{0}<|\delta|\leq\pi/2$, the ground state energy of the
system reads
\begin{eqnarray}
E &=& E_{0}+\displaystyle{E_{c}[|\Delta N|-\frac{1}{2}(\cos(2\delta)+1)]^{2}}
\nonumber \\
&-& \displaystyle{
E_{c}(\frac{2ae^{\gamma}\bar{E}_{c}\sin(2\delta)}{\hbar v_{F}})^{2}}
\label{25}\end{eqnarray}
Now we can obtain the fractional Coulomb blockade conductance peak
splitting\cite{13}. For the case of $0\leq |\delta|\leq\delta_{0}$, it can be
written as
\begin{eqnarray} 
f &=& \displaystyle{[E(\Delta N=\frac{1}{2})-E(\Delta N=0)]/
E(\Delta N=\frac{1}{2}, \delta^{c}=\pm\frac{\pi}{2})} \nonumber \\ 
&=& \displaystyle{
4(\frac{2ae^{\gamma}E_{c}\sin(2\delta)}{\hbar v_{F}})^{2}
(\frac{1+\cos(2\delta)}{2\pi})^{4}(1-\cos(\frac{\pi}{1+\cos(2\delta)}))}
\label{26}\end{eqnarray}
In the case of $\delta_{0}<|\delta|\leq\pi/2$, it can be written as
\begin{equation}
f=4[\frac{1}{2}-\frac{1}{2}(1+\cos(2\delta))]^{2}
\label{27}\end{equation}
It is noted that the fractional Coulomb blockade conductance peak splitting
significantly depends upon the phase shift $\delta$, i.e., the transmission
coeficient {\cal T}. However, in the case of the weak
backward scattering $0\leq|\delta|\leq\delta_{0}$, it also denpends upon the
parameter $a$, the Fermi velocity $v_{F}$ and the electrostatic Coulomb energy
$E_{c}=e^{2}/(2C)$. In the case of the strong backward
scattering $\delta_{0}<|\delta|\leq\pi/2$, the fractional Coulomb blockade 
conductance peak splitting $f$ only depends upon the phase shift $\delta$,
therefore, it is universal.

We now study the low temperature behavior of the tunneling conductance. To
this end, we define the following current operator
\begin{eqnarray}
I &=& \displaystyle{e\partial_{t}(U^{+}\hat{n}U)} 
=\displaystyle{e\frac{1+\cos(2\delta)}{2\pi}\partial_{t}\Phi_{-c}(0)}
\nonumber \\
&-& \displaystyle{\frac{aeD\sin(2\delta)}{\pi\hbar v_{F}}[
\partial_{t}\Phi_{-c}(0)\cos(\Phi_{-c}(0))\cos(\Phi_{-s}(0))-
\partial_{t}\Phi_{-s}(0)\sin(\Phi_{-c}(0))\sin(\Phi_{-s}(0))]}
\label{28}\end{eqnarray}
Using the Kubo formula of the conductance, we can easily obtain the following
tunneling conductance in the low energy and low temperature limit
\begin{equation}
G(T)=G_{0}(\delta)[1+AT(\frac{\sin(2\delta)}{\pi})^{2}\cos(
2<\Phi_{-c}(0)>)]
\label{29}\end{equation}
where $G_{0}(\delta)=e^{2}(1+\cos(2\delta))^{2}/(4\pi\hbar)$, 
$T$ is the temperature, and $A$ is a constant.
It is noted that the tunneling conductance has a linear temperature
dependence, and at the strong coupling critical point $\delta^{c}=\pm\pi/2$,
it is equal to zero. It is necessary to mention that as the frequency $\omega$
and the temperature $T$ tend to zero, the system approaches to the strong 
coupling critical point $delta^{c}=\pm\pi/2$ because the backward scattering
term is relevant, the renormalized backward scattering potential
$\bar{V}_{2k_{F}}$ tends to infinity in the low energy limit\cite{fisher}.  

We define the following dimensionless tunneling conductance
\begin{equation}
g=\frac{G(0)}{G_{0}}=\frac{1}{2}(1+\cos(2\delta))^{2}
\label{30}\end{equation}
where $G_{0}=e^{2}/(2\pi\hbar)$ is a unit quantum conductance of each
tunneling channel. From Equs.(\ref{26}) and (\ref{27}), we can obtain the 
following relation between the
dimensionless tunneling conductance $g$ and the fractional Coulomb blockade
peak splitting $f$
\begin{equation}
f=\left\{\begin{array}{ll}
\displaystyle{8\sqrt{2}Bg^{5/2} (1-\sqrt{\frac{g}{2}})(1-\cos(\frac{\pi}{
\sqrt{2g}}))}, & \;\;\; 0\leq |\delta|\leq\delta_{0}\\
\displaystyle{4(\frac{1}{2}-\sqrt{\frac{g}{2}})^{2}}, & \;\;\; 
\delta_{0}<|\delta|\leq\pi/2
\end{array}\right.
\label{31}\end{equation}
where $B=[ae^{\gamma}E_{c}/(\pi^{2}\hbar v_{F})]^{2}$. 
It qualitatively agrees with the experimental data in Ref.\cite{4,5}. 
Because the
factor $a$ depends upon the structure of the quantum point contact, and the
electrostatic energy $E_{c}$ is determined by the structure of the quantum
dot, the fractional Coulomb blockade peak splitting is not universal in the
weak backward scattering $0\leq |\delta |\leq\delta_{0}$. However, in the
strong backward scattering $\delta_{0}<|\delta |\leq\pi/2$, it is universal.

In summary, by using the bosonization method and the unitary transformation,
we have shown that the tunneling conductance between two quantum dots has the
linear temperature dependence in the low energy and low temperature
limit, and given a general relation between the dimensionless tunneling
conductance $g$ and the fractional Coulomb blockade peak splitting $f$ which
is valid both for the weak and strong backward scattering of 
the electrons on the impurity. Our treatment of this kind of system is 
simple and effective,
and can be used to study other similar problems because it can exactly treat
the backward scattering potential which is a relevant term in the low energy
region, the usual perturbation expansion methods of this term may fail because
the high orders are divergent in the low energy limit.

We are very grateful to Prof. Peter Fulde for his 
encouragement.

\end{document}